\documentclass[aps,showpacs,nofootinbib,prd,twocolumn]{revtex4-1}
\usepackage{graphicx}
\usepackage{amssymb}
\usepackage{amsmath}
\usepackage{xcolor}
\usepackage{mathtools,slashed}
\usepackage{epstopdf}
\usepackage[utf8]{inputenc}
\usepackage{url}
\usepackage{subfigure}
\usepackage{soul}
\usepackage{float}
\usepackage{newtxmath}

\usepackage{epsfig}

\newcommand{\be}{\begin{eqnarray}}
\newcommand{\ee}{\end{eqnarray}}
\begin{document}

\title{Collision energy dependence of the critical end point from baryon number fluctuations in the Linear Sigma Model with quarks}


\author{Alejandro Ayala$^{1,2}$}
\author{Bilgai Almeida Zamora$^3$}
\author{J. J. Cobos-Mart\'inez$^4$}
\author{S. Hern\'andez-Ortiz$^5$}
\author{L. A. Hernández$^{6,2,7}$}
\author{Alfredo Raya$^{8,9}$}
\author{Mar\'ia Elena Tejeda-Yeomans$^{10}$}
  \address{
  $^1$Instituto de Ciencias
  Nucleares, Universidad Nacional Aut\'onoma de M\'exico, Apartado
  Postal 70-543, CdMx 04510,
  Mexico.\\
  $^2$Centre for Theoretical and Mathematical Physics, and Department of Physics,
  University of Cape Town, Rondebosch 7700, South Africa.\\
  $^3$Departamento de Investigaci\'on en F\'isica, Universidad de Sonora, Boulevard Luis Encinas J. y Rosales, 83000, Hermosillo, Sonora, Mexico.\\
  $^4$Departamento de F\'isica, Universidad de Sonora, Boulevard Luis Encinas J. y Rosales, 83000, Hermosillo, Sonora, Mexico.\\
  $^5$Institute for Nuclear Theory, University of Washington, Seattle, WA, 98195, USA.\\
  $^6$Departamento de F\'isica, Universidad Aut\'onoma Metropolitana-Iztapalapa, Av. San Rafael Atlixco 186, C.P, CdMx 09340, Mexico.\\
  $^7$ Facultad de Ciencias de la Educaci\'on, Universidad Aut\'onoma de Tlaxcala, Tlaxcala, 90000, Mexico. \\
  $^8$Instituto de F\'isica y Matem\'aticas, Universidad Michoacana de San Nicol\'as de Hidalgo, Edificio C-3, Ciudad Universitaria, Francisco J. M\'ujica s/n Col. Fel\'icitas del R\'io. C. P. 58040, Morelia, Michoac\'an, Mexico.\\
  $^9$Centro de Ciencias Exactas, Universidad del Bío-Bío. Avda. Andrés Bello 720, Casilla 447, 3800708, Chillán, Chile.\\
  $^{10}$Facultad de Ciencias - CUICBAS, Universidad de Colima, Bernal D\'iaz del Castillo No. 340, Col. Villas San Sebasti\'an, 28045 Colima, Mexico.
  }
\begin{abstract}
We show that the Linear Sigma Model with quarks produces an effective description of the QCD phase diagram and of the system's equilibrium distribution properties that deviate from those of the Hadron Resonance Gas Model. The deviation is due to the inclusion of plasma screening properties, encoded in the contribution of the ring diagrams and thus to the introduction of a key feature of plasmas near phase transitions, namely, long-range correlations. After fixing the model parameters using input from LQCD for the crossover transition at vanishing chemical potential, we study the location of the Critical End Point in the effective QCD phase diagram. We use the model to study baryon number fluctuations and show that in heavy-ion collisions, the CEP can be located for collision energies $\sqrt{s_{NN}}\sim 2$ GeV, namely, in the lowest NICA or within the HADES energy domain.
\end{abstract}
\maketitle

\section{Introduction}

The phase structure of strongly interacting matter in the temperature and baryon density plane, has become a subject with growing attention over the last years. The interest is driven by the possibility to experimentally probe this phase structure by means of relativistic heavy-ion collisions using either current experimental facilities, in particular the RHIC-STAR detector with its Beam Energy Scan program and HADES, which already have shown results on net-proton number fluctuations~\cite{STAR:2020tga,HADES:2020wpc}, or dedicated facilities soon to enter into operation, such as NICA~\cite{Kekelidze:2017tgp} and FAIR~\cite{Senger:2017nvf}. The subject has received a boost in interest after the successful detection of gravitational waves from neutron star mergers. In fact, simulations show that these mergers probe similar regions in the phase diagram as those that can be studied in relativistic heavy-ion collisions~\cite{Blacker:2020nlq,Most:2019onn}. 

Calculations using Lattice QCD (LQCD) have found that for finite temperature $T$ and vanishing baryon chemical potential $\mu_B$, the transition between the confined/chiral symmetry broken and the deconfined/partially restored chirally symmetric phases, is a crossover that happens at a pseudocritical temperature $T_c(\mu_B=0)\simeq 158$ MeV~\cite{Borsanyi:2020fev,Bazavov:2018mes,Aoki:2006we}. However, using effective model calculations~\cite{Roessner:2006xn,Ayala:2019skg, Asakawa:1989bq,Ayala:2017ucc,Gao:2016qkh,Gao:2020fbl} it can be shown that this transition becomes first order for low $T$ and high $\mu_B$. Thus, when the temperature is increased, the end of this first order phase transition line in the $T$ vs.~$\mu_B$ plane should happen at a Critical End Point (CEP). Unfortunately, LQCD calculations cannot be used to directly determine the position of this CEP due to the severe sign problem~\cite{Ding:2015ona}, although early attempts where made some time ago in Ref.~\cite{Fodor:2004nz}. More recently, results employing the Taylor series expansion around $\mu_B=0$ or the extrapolation from imaginary to real $\mu_B$ values, suggest that the CEP has not yet been found for $\mu_B/T\leq 2$ and $145\leq T\leq 155$ MeV~\cite{Sharma:2017jwb,Borsanyi:2020fev}. 

Microscopic model calculations indicate that an energy density of 0.5~GeV/fm$^3$ is able to produce a transition from the hadronic phase to the phase composed of deconfined quarks and gluons and that this density can be achieved in the center of the fireball created in head-on collisions of heavy-ions at energies above $\sqrt{s_{NN}}=3-5$~GeV~\cite{Mendenhall:2020fil}. Since for low collision energies, nuclear stopping also produces a high baryon density in the interaction region, it is expected that collisions at such energies could find signals related to the appearance of the CEP. 

The strategy to locate the CEP is based on finding deviations from the statistical behavior of a gas made out of resonances, which is known to also describe the crossover transition line found by LQCD at low values of $\mu_B$~\cite{Karsch:2016yzt}. A simple description of the transition line based on resonance degrees of freedom is provided by the Hadron Resonance Gas Model (HRGM), when the occupation numbers are given in terms of Boltzmann statistics~\cite{Braun-Munzinger:2003pwq}. Recall that the statistical properties of a thermal system can be described in terms of the cumulants of its conserved charges, which are extensive quantities~\cite{Asakawa:2015ybt}. To avoid uncertainties introduced by volume effects, the analyses consider ratios of these cumulants. The strategy is therefore linked to finding deviations in the ratios of these cumulants from those obtained in the HRGM, which are described by the Skellam distribution where, in particular, the ratios of cumulants of even order are found to be equal to 1. Of the conserved quantities that can be experimentally monitored, baryon number is accessible through the measurement of proton multiplicities~\cite{STAR:2020ddh,STAR:2021iop,STAR:2021rls,HADES:2020wpc}. The search for the location of the CEP is thus guided by the expected emergence of critical behaviour~\cite{Hatta:2002sj, Stephanov:1999zu,Bzdak:2016sxg,Bzdak:2019pkr,Athanasiou:2010kw,Mroczek:2020rpm} on this and other conserved charges such as electric charge and strangeness, when a collision energy scan is performed. Recent analyses of the possible location of the CEP, that also resort to computing high order cumulants, include techniques such as Dyson-Schwinger equations~\cite{Isserstedt:2019pgx}, a QCD assisted effective model~\cite{Fu:2021oaw} and holographic methods~\cite{Grefa:2021qvt}.

In this work we study the evolution of cumulant ratios describing baryon number fluctuations as a function of the collision energy, in the context of relativistic heavy-ion collisions, using the Linear Sigma Model with quarks (LSMq). The model incorporates a crucial feature of plasmas near transition lines, namely, the plasma screening effects. These are encoded in the contribution of the ring diagrams into the effective potential which becomes a function of the order parameter after spontaneous chiral symmetry breaking. Therefore, the statistical properties of the system can be formulated in terms of fluctuations of this order parameter~\cite{Stephanov:2008qz,Stephanov:2011pb} when the collision energy $\sqrt{s_{NN}}$ and thus $T$ and $\mu_B$, are varied. We expand on the findings of Refs.~\cite{Ayala:2019skg,Ayala:2017ucc,Ayala:2015hba} where the analysis was based only on the properties of the effective potential, also up to ring diagrams order, without studying the higher order fluctuations predicted by the model and where no attempt to locate the position of the CEP as a function of the collision energy was made. In order to gain analytical insight, we resort to the high temperature approximation and work in the chiral limit. Although these approximations have limitations concerning the model accuracy for the location of the CEP, they constitute a useful guide to carry out future, more precise studies. The work is organized as follows: In Sec.~\ref{secII} we describe the LSMq and compute the effective potential up to the ring diagrams contribution, which requires computation of the self-energies corresponding to the meson degrees of freedom. We fix the model parameters requiring that the phase transition line near $\mu_B=0$ corresponds to the one found by the latest LQCD calculations~\cite{Borsanyi:2020fev}. In Sec.~\ref{secIII} we formulate the way the baryon number fluctuations can be described in terms of the probability distribution associated to the order parameter near the transition line and present the results of the analysis for the ratios of cumulants describing baryon number fluctuations as a function of $\sqrt{s_{NN}}$, showing that these ratios deviate from the expectations of the HRGM for energies around $\sqrt{s_{NN}}\sim 4-6$ GeV and that the CEP can be found at energies $\sqrt{s_{NN}}\sim 2$ GeV. The model thus predicts that the CEP can be found either in the lowest NICA or within the HADES energy domain. We finally summarize and conclude in Sec.~\ref{concl}.

\section{The Linear Sigma Model with quarks}\label{secII}

Effective models are useful to identify the
main characteristics of the QCD phase diagram. Although no single model can be used to describe the whole phase diagram, they can be employed to explore different regions when including different degrees of freedom. For instance, the Polyakov loop extended Nambu--Jona-Lasinio or quark-meson models have been employed in Refs.~\cite{Kashiwa:2007hw,Mao:2009aq,Skokov:2010wb,Skokov:2010uh} to study simultaneously the deconfinement and chiral symmetry restoration aspects of QCD. Given that LQCD calculations, extended to small values of $\mu_B$, find coincident transition lines for the deconfinement and chiral symmetry restoration transitions, it should be possible to explore
the phase diagram emphasizing independently either
the deconfinement or the chiral aspects of the transition. 

An effective model based on chiral symmetry is provided by the LSMq. The Lagrangian is given by 
\begin{eqnarray}
   \!\!\!\!\!\!\mathcal{L}&=&\frac{1}{2}(\partial_\mu \sigma)^2  + \frac{1}{2}(\partial_\mu \vec{\pi})^2 + \frac{a^2}{2} (\sigma^2 + \vec{\pi}^2)\nonumber\\
   &-& \frac{\lambda}{4} (\sigma^2 + \vec{\pi}^2)^2 + i \bar{\psi} \gamma^\mu \partial_\mu\psi -g\bar{\psi} (\sigma + i \gamma_5 \vec{\tau} \cdot \vec{\pi} )\psi ,
\label{lagrangian}
\end{eqnarray}
where $\psi$ is an $SU(2)$ isospin doublet of quarks,
\begin{eqnarray}
 \vec{\pi}=(\pi_1, \pi_2, \pi_3 ),
\end{eqnarray} and $\sigma$ are an isospin triplet and singlet, corresponding to the three pions and the sigma, respectively. The squared mass parameter $a^2$ and the self-coupling $\lambda$ and $g$ are taken to be positive and, for the purpose of describing the chiral phase transition at finite $T$ and $\mu_B$, they need to be determined from conditions close to the phase boundary, and not from vacuum conditions. The original LSM, was introduced to describe the interaction between pions nucleons in the seminal work of Ref.~\cite{Gell-Mann:1960mvl}. A review of the LSM at finite temperature is provided in Ref.~\cite{Petropoulos:1998gt}. An early study of the phase diagram using LSMq is provided in Ref.~\cite{Bowman:2008kc}.

Working in the strict chiral limit, to allow for a spontaneous symmetry breaking, we let the $\sigma$ field to develop a vacuum expectation value $v$, namely
\begin{eqnarray} \sigma \rightarrow \sigma + v.
\label{shift}
\end{eqnarray}
This vacuum expectation value can be identified with the order parameter of the theory. After this shift, the Lagrangian can be rewritten as
\begin{eqnarray}
{\mathcal{L}}&=& \frac{1}{2}(\partial_\mu \sigma)^2  + \frac{1}{2}(\partial_\mu \vec{\pi})^2-\frac{1}
   {2}\left(3\lambda v^{2}-a^{2} \right)\sigma^{2}\nonumber\\
   &-&\frac{1}{2}\left(\lambda v^{2}- a^2 \right)\vec{\pi}^{2}\nonumber\\
   &+&\frac{a^{2}}{2}v^{2} -\frac{\lambda}{4}v^{4} + i \bar{\psi} \gamma^\mu \partial_\mu\psi
  -gv \bar{\psi}\psi + {\mathcal{L}}_{I}^b + {\mathcal{L}}_{I}^f,
  \label{lagranreal}
\end{eqnarray}
where ${\mathcal{L}}_{I}^b$ and  ${\mathcal{L}}_{I}^f$ are given by
\begin{eqnarray}
  {\mathcal{L}}_{I}^b&=&-\frac{\lambda}{4}\Big[(\sigma^2 + (\pi^0)^2)^2+ 4\pi^+\pi^-(\sigma^2 + (\pi^0)^2 + \pi^+\pi^-)\Big],\nonumber\\
  {\mathcal{L}}_{I}^f&=&-g\bar{\psi} (\sigma + i \gamma_5 \vec{\tau} \cdot \vec{\pi} )\psi.
  \label{lagranint}
\end{eqnarray}

The terms given in Eq.~(\ref{lagranint}) describe the interactions among the fields $\sigma$, $\vec{\pi}$ and $\psi$, after symmetry breaking. Since no external agent that could couple to the electric charge, such as a magnetic field, is considered, there is no distinction between neutral and charged pions. From Eq.~(\ref{lagranreal}) one can see that the $\sigma$, the three pions and the quarks have masses given, respectively, by
\begin{eqnarray}
  m^{2}_{\sigma}&=&3  \lambda v^{2}-a^{2},\nonumber\\
  m^{2}_{\pi}&=&\lambda v^{2}-a^{2}, \nonumber\\
  m_{f}&=& gv.
\label{masses}
\end{eqnarray}
Notice that the square of the tree-level boson masses in Eq.~(\ref{masses}) can vanish or even become negative as the continuous order parameter $v$ spans its domain. The physical masses, including their thermal part, are computed when $v$ takes on the value found from minimizing the effective potential. In this work we have not focused explicitly on describing the behavior of these thermal masses. Nevertheless, our results are perfectly consistent with results of previous works such as the one in Ref.~\cite{Scavenius:2000qd} when including the thermal contribution to these masses, which are provided by their self-energies (see Eq.~(\ref{Pileading}) below), albeit, in our case, in the high temperature limit.

In order to determine the chiral symmetry restoration conditions as function of $T$ and $\mu_B$, we study the behavior of the effective potential which includes the classical potential or tree-level contribution, the one-loop correction both for bosons and fermions and the ring diagrams contribution, which accounts for the plasma screening effects. 

The tree-level potential is given by
\begin{equation}
    V^{\text{tree}}(v)=-\frac{a^2}{2}v^2+\frac{\lambda}{4}v^4,
    \label{treelevel}
\end{equation}
whose minimum is found at
\begin{equation}
    v_0=\sqrt{\frac{a^2}{\lambda}}.
\end{equation}
Since $v_0\neq 0$, we notice that chiral symmetry is spontaneously broken. However, in order to make sure that the quantum corrections at finite temperature and density maintain the general properties of the effective potential, we need to add counter-terms $\delta a^2$ and $\delta \lambda$ to the bare constants $a^2$ and $\lambda$, respectively, and write \begin{align}
	V^{\text{tree}}&=-\frac{a^2}{2}v^2+\frac{\lambda}{4}v^4\rightarrow \nonumber \\
    V^{\text{vac}}&= -\frac{(a^2+\delta a^2)}{2}v^2+\frac{(\lambda+\delta \lambda)}{4}v^4+V^{\text{1;vac}},
    \label{newtree}
\end{align}
where $V^{\text{1;vac}}$ is the boson and fermion one-loop vacuum contribution. These counterterms need to be fixed from the conditions to keep the vacuum and its curvature fixed at their tree-value levels, namely,
\begin{eqnarray}
\frac{1}{v}\frac{dV^{\text{vac}}}{dv}\Big|_{v=v_0}=0,\nonumber\\
\frac{d^2V^{\text{vac}}}{dv^2}\Big|_{v=v_0}=2a^2.
\label{stability}
\end{eqnarray}
The procedure is dubbed {\it vacuum stabilization}.
It can be shown~\cite{Ayala:2021nhx} that after this procedure is implemented, the vacuum potential $V^{\text{vac}}$ coincides with tree level potential $V^{\text{tree}}$ near the minimum and that $V^{\text{vac}}$ is independent of the chosen ultraviolet scale. Therefore, when working near the minimum, as is the case of this work, in order to avoid writing long analytical expressions, we keep working with the vacuum potential at tree level which is the practical result after implementing vacuum stabilization.

To include quantum corrections at finite temperature and baryon density, we work within the imaginary-time formalism of thermal field theory. The general expression for the one-loop boson contribution can be written as
\begin{equation}
    V^{\text{b}}(v,T)=T\sum_n\int\frac{d^3k}{(2\pi)^3} \ln D_{\text{b}}(\omega_n,\vec{k})^{1/2},
    \label{1loopboson}
\end{equation}
where
\begin{equation}
D_\text{b}(\omega_n,\vec{k})=\frac{1}{\omega_n^2+k^2+m_b^2},
\end{equation} 
is the free boson propagator with $m_b^2$ being the square of the boson mass and $\omega_n=2n\pi T$ the Matsubara frequencies for boson fields. 

For a fermion field with mass $m_f$, the general expression for the one-loop correction at finite temperature and quark chemical potential $\mu=\mu_B/3$ is
\begin{equation}
    V^{\text{f}}(v,T,\mu)=-T\sum_n\int\frac{d^3k}{(2\pi)^3} \text{Tr}[\ln S_\text{f}(\tilde{\omega}_n,\vec{k})^{-1}],
    \label{1loopfermion}
\end{equation}
where
\begin{equation}
S_\text{f}(\tilde{\omega}_n,\vec{k})=\frac{1}{\gamma_0( \tilde{\omega}_n+i\mu)+\slashed{k}+m_f},
\end{equation}
is the free fermion propagator and $\tilde{\omega}_n=(2n+1)\pi T$ are the Matsubara frequencies for fermion fields. 

The ring diagrams term is given by
\begin{eqnarray}
    \!\!\!\!\!\!\!\!\!\!\!V^{\text{Ring}}(v,T,\mu)=\frac{T}{2}\sum_n\int\frac{d^3k}{(2\pi)^3}\ln [1+\Pi_{\text{b}}D_b(\omega_n,\vec{k})],
    \label{rings}
\end{eqnarray}
where $\Pi_\text{b}$ is the boson self-energy. The self-energies for the sigma and pion fields are in general different. However, in this work we resort to the high-temperature approximation, keeping only the leading matter effects. In this approximation the boson self-energies become independent of the boson species and we can write~\cite{Ayala:2021nhx}
\begin{eqnarray}
\!\!\!\!\!\Pi_{\text{b}}&\equiv&\Pi_\sigma=\Pi_{\pi^\pm}=\Pi_{\pi^0}\nonumber\\
\!\!\!\!\!&=&\lambda\frac{T^2}{2}-N_\text{f}N_\text{c}g^2\frac{T^2}{\pi^2}\left[\text{Li}_2\left(-e^{-\frac{\mu}{T}}\right)+\text{Li}_2\left(-e^{\frac{\mu}{T}}\right)\right],
\label{Pileading}
\end{eqnarray}
where $N_\text{c}$ and $N_\text{f}$ are the number of colors and flavors, respectively, whereas ${\rm Li}_2(x)$ stands for the polylogarithm function of order 2 (dilogarithm).

Writing all the ingredients together, the finite temperature and baryon density effective potential, up to the ring diagrams contribution, after vacuum stabilization and boson and fermion mass renormalizaton in the $\overline{\text{MS}}$ scheme at the ultraviolet scale $\tilde{\mu}$, can be written as
\begin{widetext}
\begin{eqnarray}
V^{\text{eff}}(v)&=&-\frac{a^{2}}{2}v^{2}+\frac{\lambda}{4}v^{4}+\sum_{\text{b}=\pi^\pm,\pi^0,\sigma}\left\{-\frac{T^4\pi^2}{90}+\frac{T^2m_\text{b}^2}{24}-\frac{T(m_\text{b}^2+\Pi_\text{b})^{3/2}}{12\pi}-\frac{m_\text{b}^4}{64\pi^2}\left[\ln\left(\frac{\tilde{\mu}^2}{(4\pi T)^2}\right)+2\gamma_E\right]\right\}\nonumber\\
&+&N_\text{c}N_\text{f}\left\{\frac{m_f^4}{16\pi^2}\Big[\ln\left(\frac{\tilde{\mu}^2}{T^2}\right)- \psi^{0}\left(\frac{1}{2}+\frac{i\mu}{2\pi T}\right)- \psi^{0}\left(\frac{1}{2}-\frac{i\mu}{2\pi T}\right)\right.+
\psi^{0}\left(\frac{3}{2}\right)-2\left(1+\ln(2\pi)\right)+\gamma_E\Big]\nonumber\\
&-&\left.\frac{m_\text{f}^2T^2}{2\pi^2}\left[\text{Li}_2\left(-e^{-\frac{\mu}{T}}\right)+\text{Li}_2\left(-e^{\frac{\mu}{T}}\right)\right]+\frac{T^4}{\pi^2}\left[\text{Li}_4\left(-e^{-\frac{\mu}{T}}\right)+\text{Li}_4\left(-e^{\frac{\mu}{T}}\right)\right]\right\},
\label{vefftot}
\end{eqnarray}
\end{widetext}
$\gamma_E\simeq 0.57721$ denoting the Euler-Mascheroni constant.
Equation~(\ref{vefftot}), with the boson self-energies given in Eq.~(\ref{Pileading}), constitute the main tools to study the effective QCD phase diagram from the point of view of chiral symmetry restoration/breaking. Notice that inclusion of the plasma screening effects, encoded in the contribution of the ring diagrams, produces the possibility that the effective potential develops a cubic term in the order parameter $v$. This happens for values of $T$ and $\mu_B$ such that $\Pi_\text{b}-a^2\simeq0$. This signals the possibility of a first order phase transition and of fluctuations that deviate from those of a free gas of mesons and quarks for those values of $T$ and $\mu_B$. 
Recall that the boson self-energy represents the matter correction to the boson mass. For a second order (our proxy for a crossover), namely, a continuous phase transition, these corrections should produce that the thermal boson masses vanish when the symmetry is restored. This means that at the phase transition, the effective potential not only develops a minimum but it is also flat (the second derivative vanishes) at $v=0$. This property can be exploited to find the suitable values of the model parameters $a$, $\lambda$ and $g$ at the critical temperature $T_c^0$ for $\mu_B^c=0$. 
Since the thermal boson masses are degenerate at $v=0$, the condition to produce a flat effective potential at $T_c$ for $\mu_B=0$ 
can be written as~\cite{Ayala:2021nhx}
\begin{eqnarray}
6&\lambda&\left(\frac{T_c^2}{12}-\frac{T_c}{4\pi}\left(\Pi_{\text{b}}-a^2\right)^{1/2}\right. \nonumber\\
&+&\left. \frac{a^2}{16\pi^2}\left[\ln\left(\frac{\tilde{\mu}^2}{(4\pi T_c)^2}\right)+2\gamma_E\right]\right)\nonumber\\
&+&g^2T_c^2-a^2=0.
\label{condatTc}
\end{eqnarray}
where, from Eq.~(\ref{Pileading}),
\begin{eqnarray}
\Pi_{\text{b}}(T_c^0,\mu_B^c=0)=\left[\frac{\lambda}{2} + g^2\right](T_c^0)^2.
\label{Pi1formu0}
\end{eqnarray}

Given that we have a relative freedom to chose the value of $\tilde{\mu}$, we take $\tilde{\mu}=500$ MeV, which is chosen large enough so as to consider it to be the largest energy scale in the problem. The dependence of the result is only logarithmic in $\tilde{\mu}$, therefore small variations in this parameter do not affect significantly the result. We look for solutions of Eq.~(\ref{condatTc}) for the set of parameters $\lambda$, $g$ and $a$, using input form the LQCD transition curve at $T_c^0\simeq 158$ MeV from where we take the values of the curvature parameters $\kappa_2$ and $\kappa_4$. Since Eq.~(\ref{condatTc}) is non-linear, the set of parameters is not unique. We work with the parametrization of the LQCD transition curve given by
\begin{equation}
   \frac{T_c(\mu_B)}{T_c^0}=1-\kappa_2\left(\frac{\mu_B}{T_c^0}\right)^2+\kappa_4\left(\frac{\mu_B}{T_c^0}\right)^4,
    \label{paraLQCD}
\end{equation}
\begin{figure}[b]
    \centering
    \includegraphics[scale=0.56]{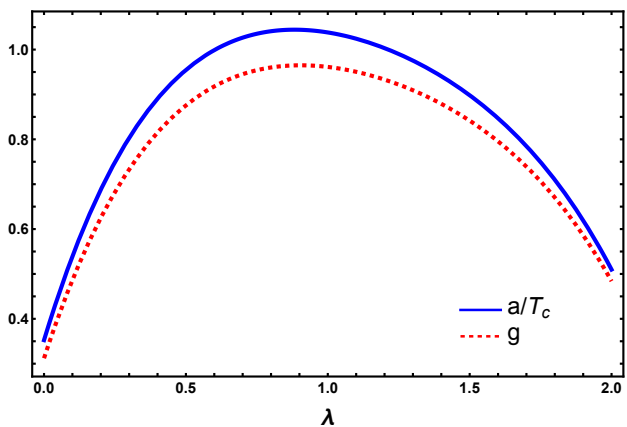}
    \caption{Parameters $g$ and $a/T_c$ as functions of $\lambda$. These are obtained from the solutions of Eq.~(\ref{condatTc}) and using the parametrization of the transition curve in Eq.~(\ref{paraLQCD}) with $\kappa_2$ and $\kappa_4$ found from the LQCD calculation in Refs.~\cite{Borsanyi:2020fev,Guenther:2020jwe}.}
    \label{ganda}
\end{figure}
with $\kappa_2=0.0153$ and $\kappa_4=0.00032$~\cite{Borsanyi:2020fev,Guenther:2020jwe}. Notice that although Eq.~(\ref{paraLQCD}) is provided in Ref.~\cite{Borsanyi:2020fev} as an implicit equation, for simplicity and in order to avoid a further layer of complexity into the analysis, we have normalized the variable $\mu_B$ to $T_c^0$. To find the values of the parameters, we first fix a value of $\lambda$, and find the solution of Eq.~(\ref{condatTc}) for $g$ and $a$ that, from the effective potential in Eq,~(\ref{vefftot}), produce a phase transition at the values of $T_c(\mu_B)$, hereafter simply referred to as $T_c$, and $\mu_B$ given by Eq.~(\ref{paraLQCD}).
We then repeat the procedure for other values of $\lambda$.  In this manner, the solutions can be expressed as a relation between the couplings $g$ and $\lambda$
\begin{equation}
    g(\lambda)=0.31+1.94 \lambda-2.06 \lambda^2+0.97 \lambda^3-0.20 \lambda^4,
\end{equation}
and a relation between $a$ and $\lambda$
\begin{eqnarray}
    \!\!\!\!\!\frac{a(\lambda)}{T_c}&=&0.35+2.08 \lambda-2.21 \lambda^2+1.03 \lambda^3-0.21 \lambda^4.
    \label{arel}
\end{eqnarray}

\begin{figure}[t]
    \centering
    \includegraphics[scale=0.5
    ]{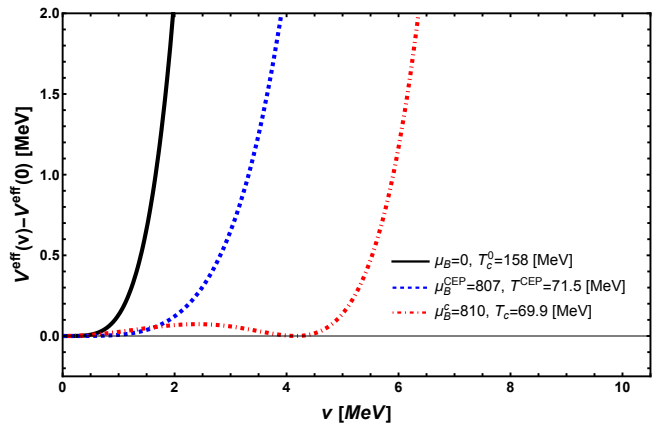}
    \caption{
    Effective potential as a function of the order parameter for three different sets of values of $T_c$ and $\mu_B^c$ along the transition curve using $a=148.73$ MeV, $\lambda=1.4$ and $g=0.88$. For $\mu_B=0$ and $T_c^0=158$ MeV the phase transition is second order. For $\mu_B^{\text{CEP}}=807$ MeV and $T^{\text{CEP}}=71.5$ MeV, where the CEP is located, the phase transition becomes first order. For $\mu_B^c> \mu_B^{\text{CEP}}$ and $T_c<T^{\text{CEP}}$, the phase transitions are always first order. This is signaled by the development of a barrier between degenerate minima at the phase transition.}
    \label{Veff}
\end{figure}
\begin{figure}[t]
        \centering
    \includegraphics[scale=0.58]{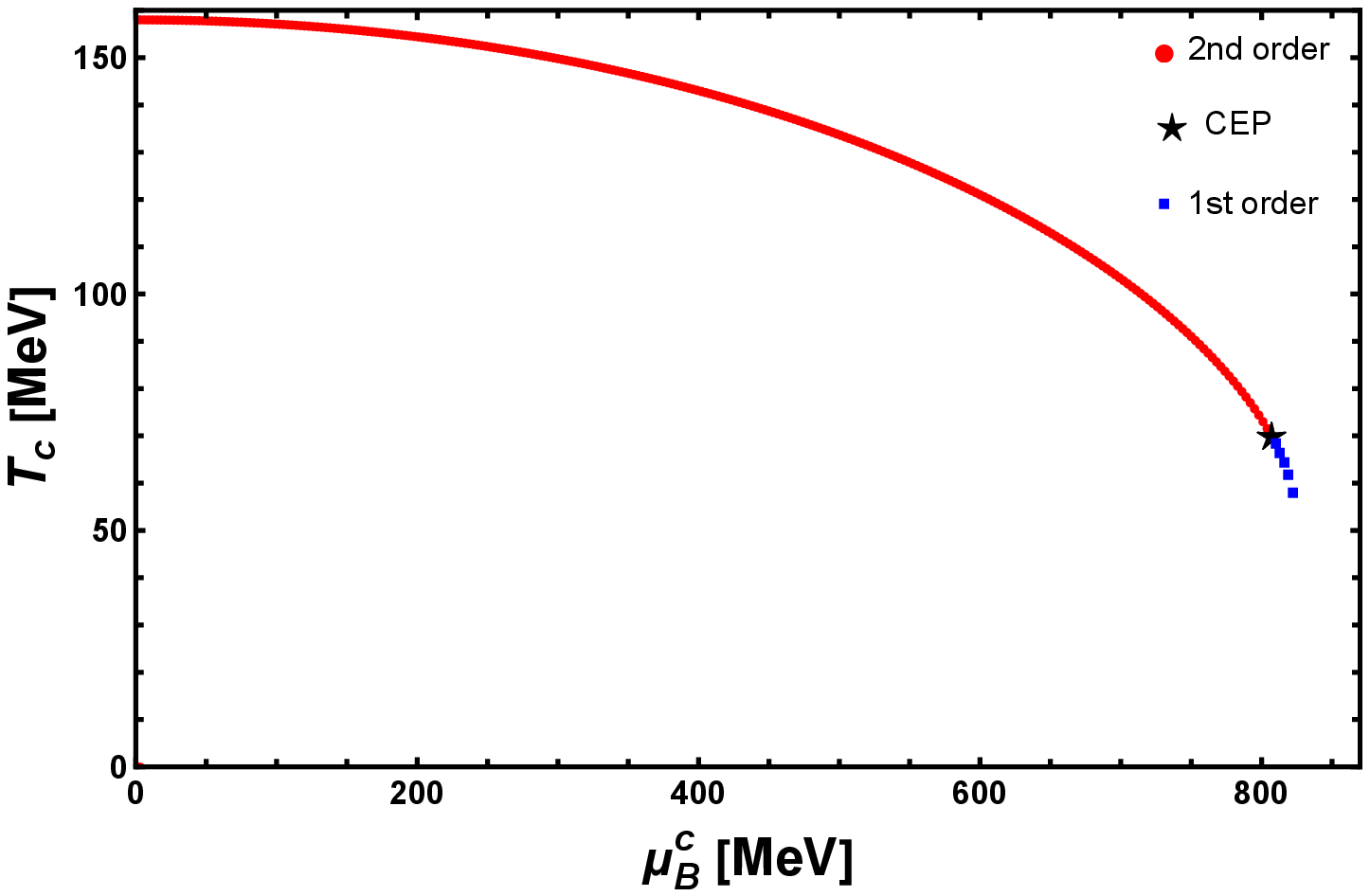}
    \\[\bigskipamount]
        \centering
    \includegraphics[scale=0.58]{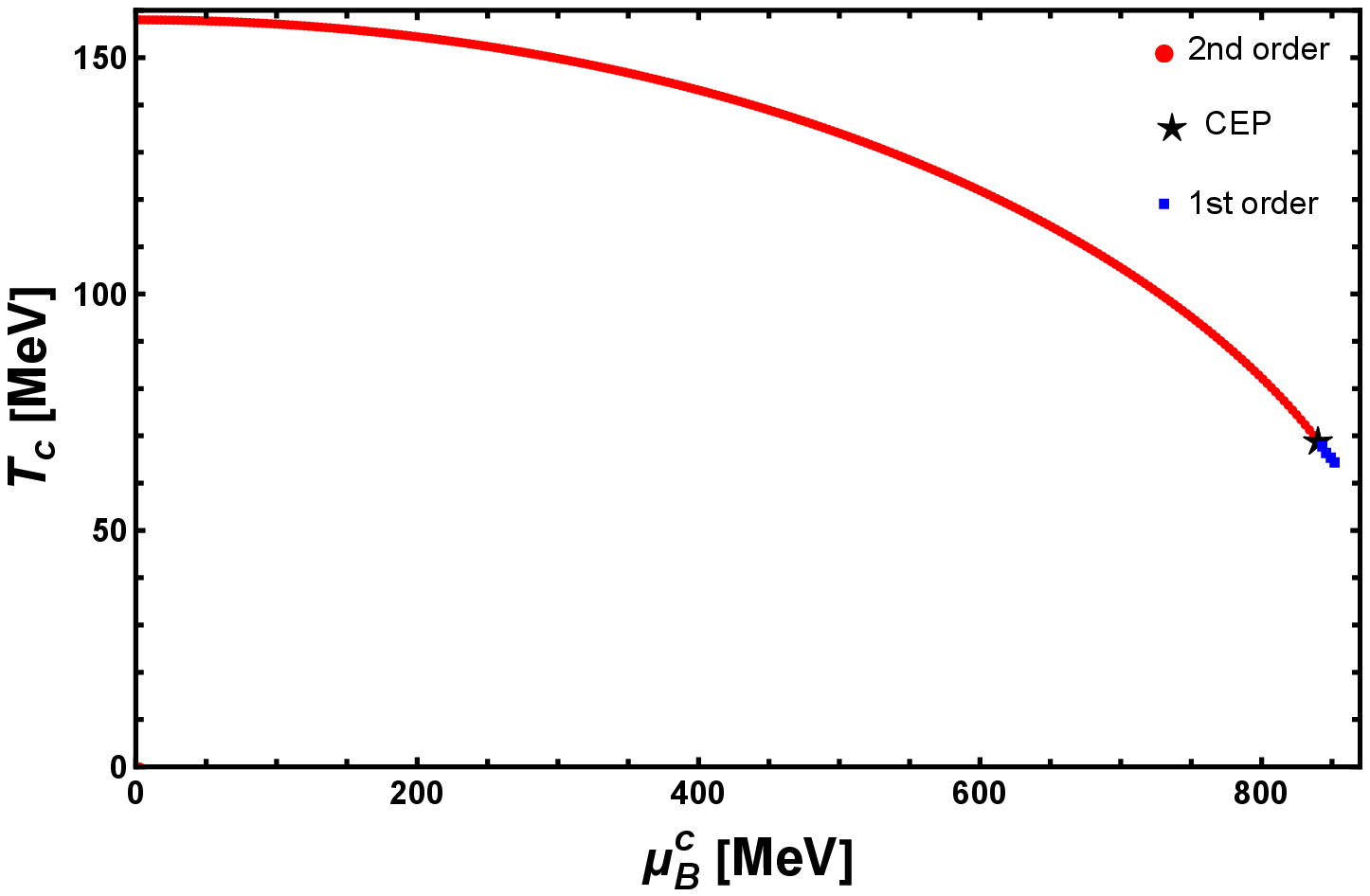}
    \caption{Examples of effective phase diagrams obtained for two choices of the possible sets of parameters $a$, $\lambda$ and $g$. The upper panel is computed with $\lambda=1.4$, $g=0.88$ and $a=148.73$ MeV. The lower panel is computed with $\lambda=0.4$, $g=0.82$ and $a=141.38$ MeV. Notice that the position of the CEP is not significantly altered by varying the choice of parameters. For the allowed range of values of $a$, $\lambda$ and $g$, the CEP location ranges between 786 MeV $<\mu_B^{\text{CEP}}<849$ MeV and 69 MeV $< T^{\text{CEP}}<70.3$ MeV.}
    \label{PD}
\end{figure}

Figure~\ref{ganda} shows the values of the parameters $g$ and $a/T_c$ for the range of allowed values of $\lambda$ where Eq.~(\ref{condatTc}) produces solutions.
With all the parameters fixed, we can study the properties of the effective potential to find the values of $T_c$ and $\mu_B^c$ where the chiral phase transition takes place. Figure~\ref{Veff} shows the effective potential as a function of the order parameter. We take as examples three different sets of values of $T_c$ and $\mu_B^c$ along the transition curve using $a=148.73$ MeV, $\lambda=1.4$ and $g=0.88$. Notice that for $\mu_B=0$ and $T_c=158$ MeV the phase transition is second order. As $\mu_B$ increases, the phase transition is signaled by a flatter effective potential until the chemical potential and temperature reach values $\mu_B^{\text{CEP}}$ and $T^{\text{CEP}}$ where the effective potential develops a barrier between degenerate minima. For $\mu_B^c>\mu_B^{\text{CEP}}$ and $T_c<T^{\text{CEP}}$, the phase transitions are always first order. These features are illustrated in Fig.~\ref{PD} where we show two examples of the effective phase diagram thus obtained. The upper panel is computed using $\lambda=1.4$, $g=0.88$ and $a=148.73$ MeV. 
The lower panel is computed with $\lambda=0.4$, $g=0.82$ and $a=141.38$ MeV. The solid (red) lines represent second order phase transitions, our proxy for crossover transitions, whereas the dotted (blue) lines correspond to first order phase transitions. The star symbol represents the location of the CEP. Notice that the CEP appears only for large values of $\mu_B$.  Also, variations of the model parameters do not change the CEP position appreciably. In fact, for the allowed range of values of $a$, $\lambda$ and $g$, the CEP location ranges between  786 MeV $<\mu_B^{\text{CEP}}<849$ MeV and 69 MeV $< T^{\text{CEP}}<70.3$ MeV.
\begin{figure}[t]
    \centering
    \includegraphics[scale=0.58
    ]{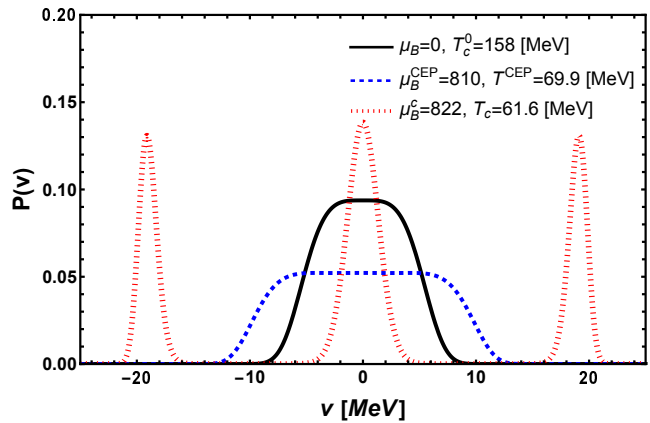}
    \caption{
    Normalized probability distribution for different pairs of $\mu_B^c$, $T_c$ along the transition curve. We use the values of the parameters $\lambda=1.4$, $g=0.88$ and $a=148.73$ MeV. For $\mu_B=0$ and $T_c^0=158$ MeV the probability distribution is Gaussian-like albeit wider. For $\mu_B^{\text{CEP}}$ and $T^{\text{CEP}}$ the probability distribution becomes even wider. For $\mu_B^c>\mu_B^{\text{CEP}}$ and $T_c<T^{\text{CEP}}$, the phase transitions are always first order and the probability distribution develops secondary peaks.
    }
    \label{Boltzmannfactor}
\end{figure}

Notice that the high temperature approximation is justified, in spite of the large values found for the $\mu_B^{\text{CEP}}$. This happens because the quantity that directly enters the calculation is the quark chemical potential $\mu$, related to $\mu_B$ by $\mu_B=3\mu$. Thus, although $\mu_B^\text{CEP}$ is rather large, $\mu^\text{CEP}$ is only $\sim 1.7\ T_c$. For this value of $\mu^\text{CEP}$, the combination of polylogarithmic functions and factors of $T$ in the second term of the right-hand side of Eq.~(\ref{Pileading}) makes that the boson masses become $\sim\sqrt{\lambda + 2g^2}\ T^{\text{CEP}}$. At the same time, $T^{\text{CEP}} \sim\ 0.5 T_c$. Thus, the large temperature approximation is still valid around $(\mu_B^\text{CEP}, T^\text{CEP})$. For even larger values of $\mu$, the absolute value of the combination of polylogarithmic functions in Eq.~(\ref{Pileading}) grows faster than the decrease of the temperature and thus the high-temperature approximation is no longer valid.

Also, notice that the quark contribution, and thus the $\mu$ dependence of the thermal boson masses, enters the calculation, according to  Eq.~(\ref{Pileading}), as the argument of an exponential function and is controlled by $g^2$. This dependence translates into the effective potential and thus into the determination of the value of $\mu_B^\text{CEP}$. Thus, variations of $\mu$ into the exponent are amplified when varying $g$ and are, accordingly, more prominent than those of $T^\text{CEP}$, which, also according to Eq.~(\ref{Pileading}), are mainly controlled by $\lambda$.

\section{Baryon number fluctuations}\label{secIII}

In order to study the fluctuations in the baryon number that the analysis predicts, we first look at the probability distribution. This is given by the expression
\begin{eqnarray}
{\mathcal{P}(v)}=\exp\left\{-\Omega V^{\text{eff}}(v)/T\right\},
\label{Boltzfac}
\end{eqnarray}
as a function of the order parameter around the equilibrium value in the symmetry restored phase $\langle v\rangle=0$. In Eq.~(\ref{Boltzfac}), the factor $\Omega$ represents the system volume. For the purposes of this work, we explore large volumes compared to the typical size of the fireball crated in heavy-ion reactions, so as to mimic the thermodynamic limit. The normalized probability distribution is illustrated in Fig.~\ref{Boltzmannfactor} for different pairs of $\mu_B^c$, $T_c$ along the transition curve using $\lambda=1.4$, $g=0.88$ and $a=148.73$ MeV. We have extended the domain of the order parameter making $v\to |v|$ to include negative values so as to ensure that its average satisfies $\langle v\rangle=0$.
Notice that for $\mu_B=0$ and $T_c^0=158$ MeV, for which the phase transition is second order, the probability distribution is Gaussian-like, albeit wider. 
\begin{figure}[t]
        \centering
    \includegraphics[scale=0.58]{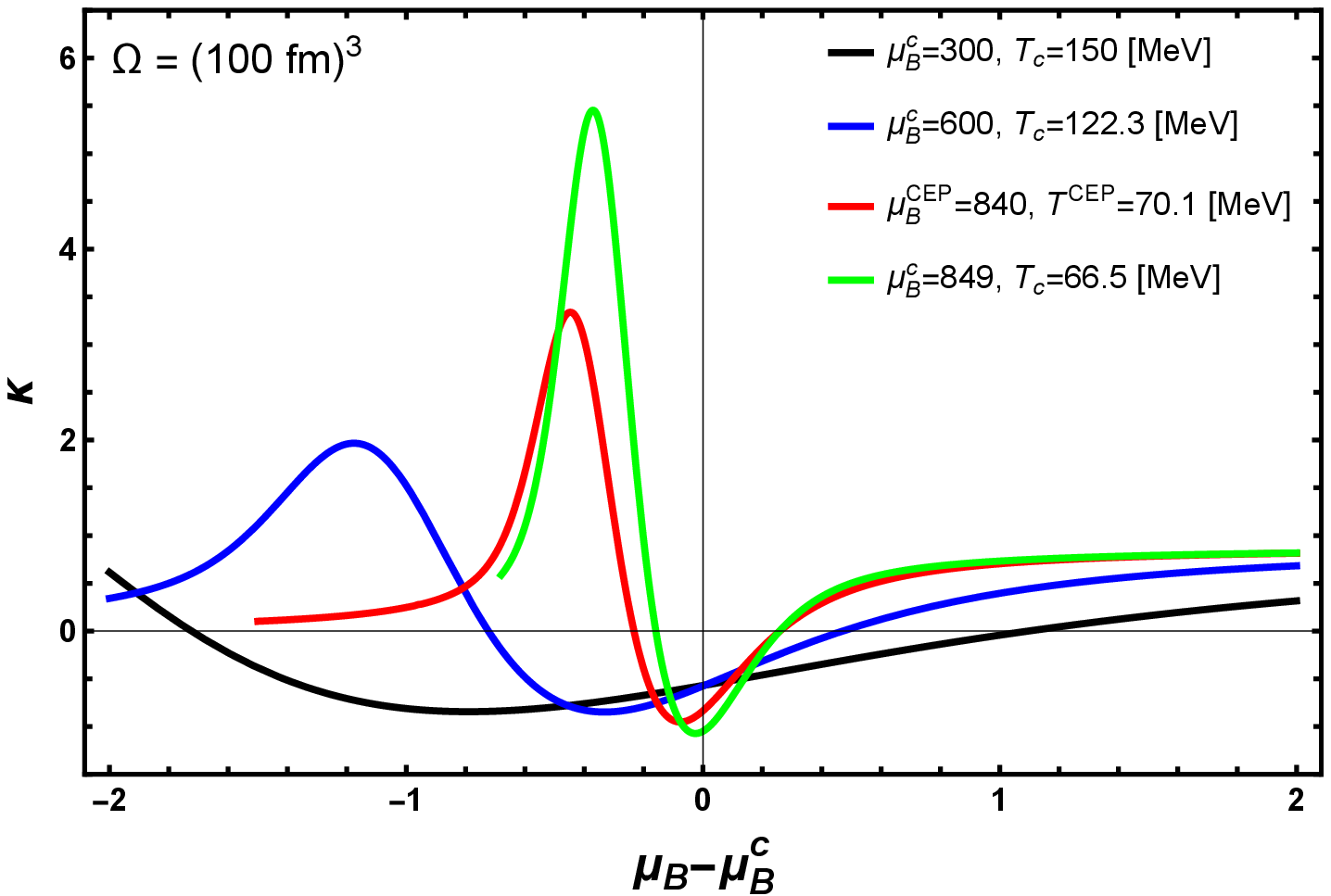}
    \\[\bigskipamount]
        \centering
    \includegraphics[scale=0.58]{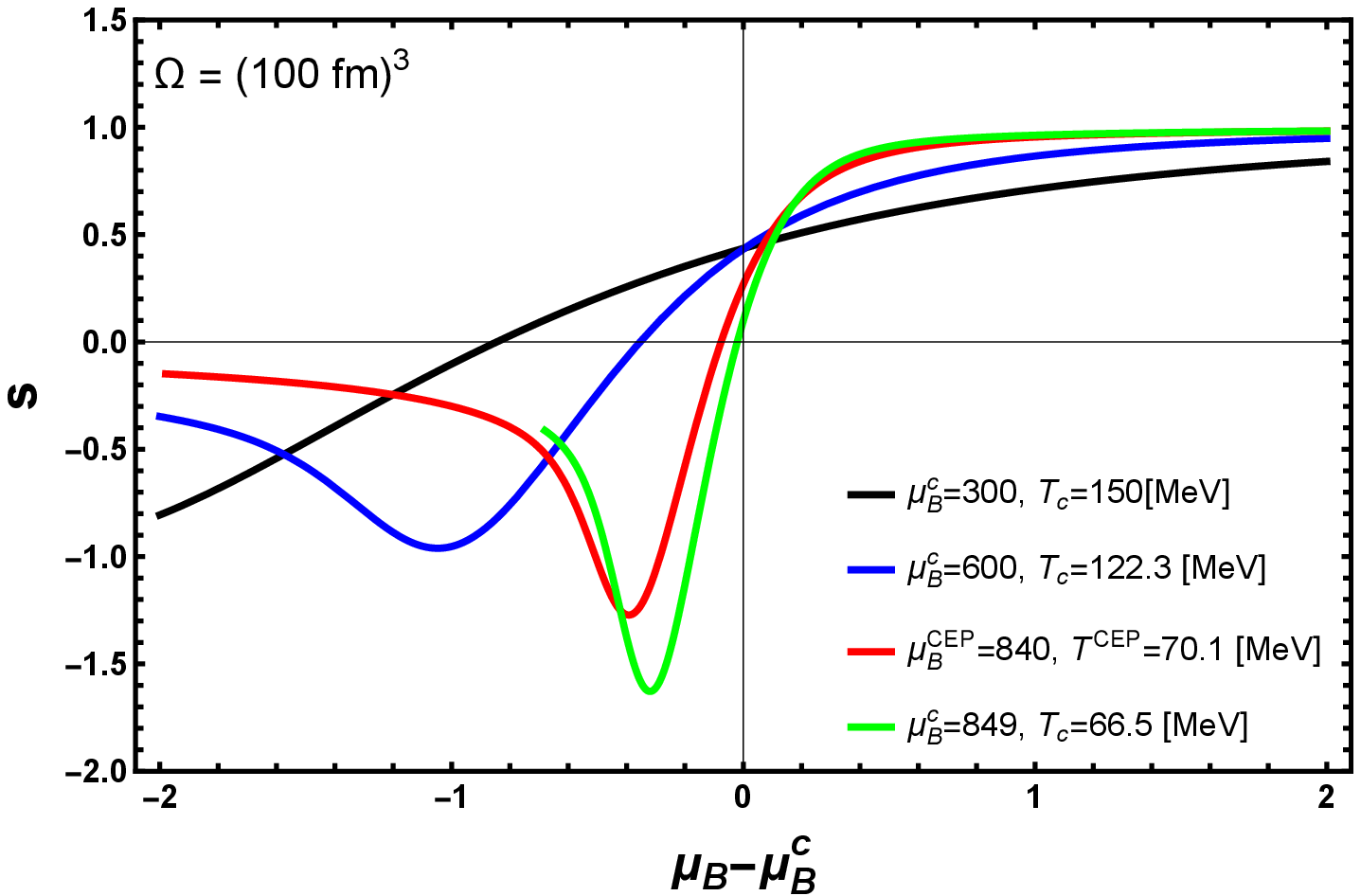}
    \caption{
    Kurtosis  $\kappa$ (upper panel) and skewness $s$ (lower panel) as functions of $\mu_B-\mu_B^c$ at a fixed value of the corresponding $T_c$ for $\lambda=0.4$, $g=0.82$ and $a=141.38$ MeV and a volume $\Omega=(100\ \text{fm})^3$. For first order phase transitions the two cumulants show a peaked structure when $\mu_B$ goes across the corresponding critical value of $\mu_B^c$.}
    \label{sandk}
\end{figure}
This happens because for these values of $\mu_B^c$ and $T_c$ the quartic terms in $v$ are important, producing a wider distribution around $\langle v\rangle$. As $\mu_B^c$ increases and the phase transition becomes first order for $\mu_B^{\text{CEP}}$ and $T^{\text{CEP}}$ the probability distribution becomes even wider. For $\mu_B^c>\mu_B^{\text{CEP}}$ and $T_c<T^{\text{CEP}}$, the phase transitions are always first order and the probability distribution develops secondary peaks that reflect the fact that for first order phase transitions the effective potential develops degenerate minima. We emphasize that the features of the probability distributions for first order phase transitions are due to the inclusion of the ring diagrams and thus of the plasma screening. Were these effects not to be included, the development of secondary peaks in the probability distribution would not happen and thus deviations from the Skellam statistics would not be possible.
\begin{figure}[t]
        \centering
    \includegraphics[scale=0.59]{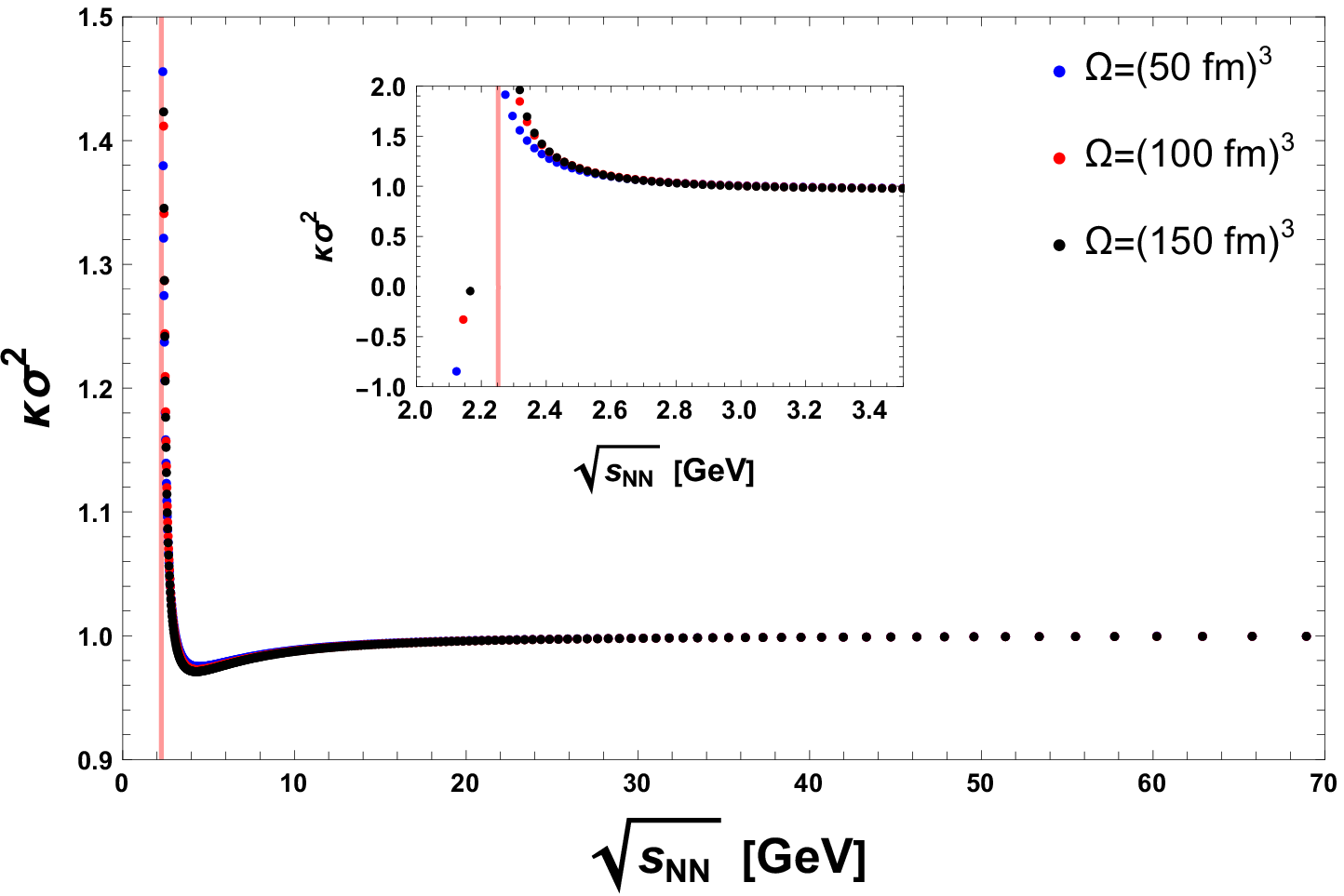}
    \\[\bigskipamount]
        \centering
    \includegraphics[scale=0.59]{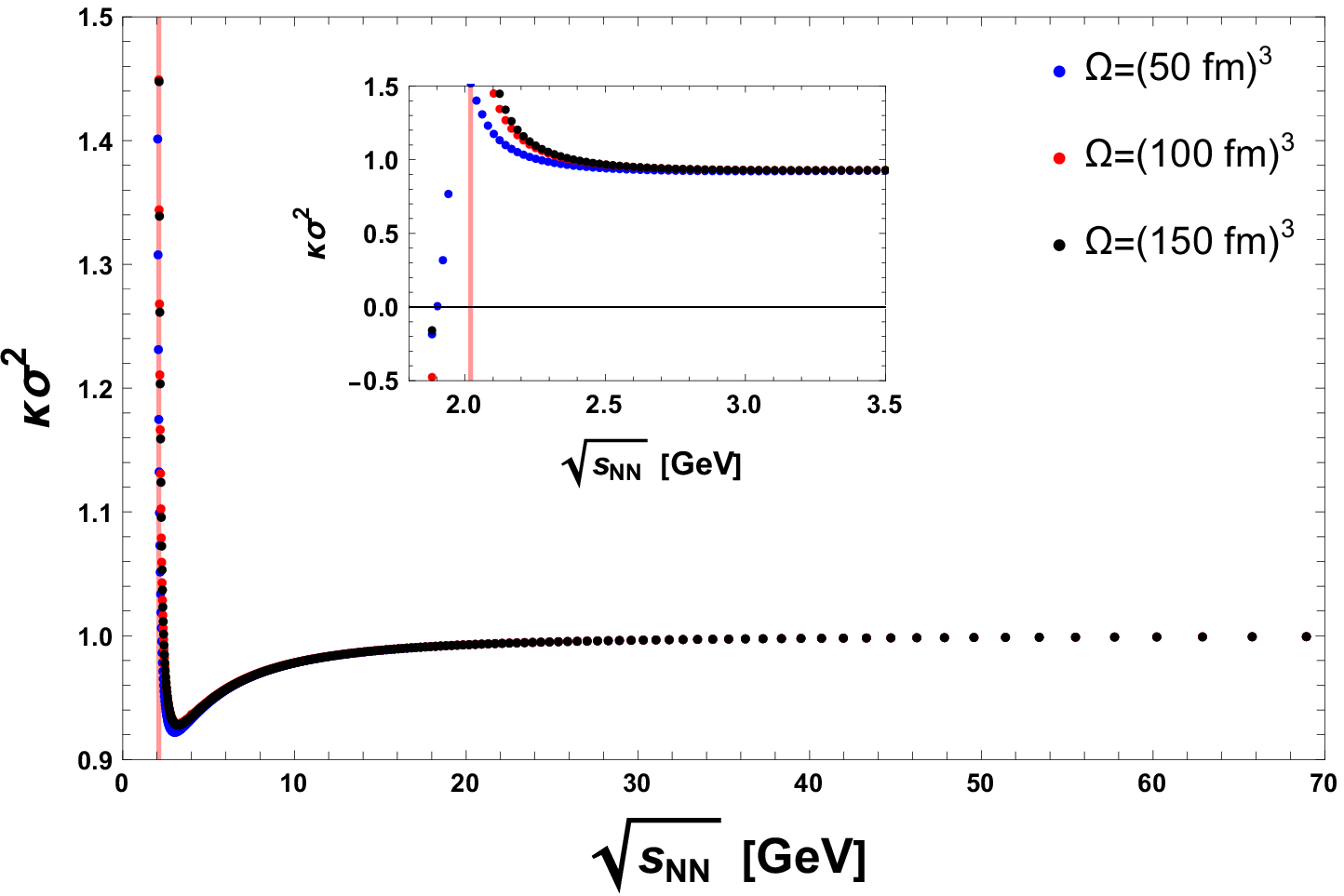}
    \caption{Ratio of the cumulants $C_4/C_2=\kappa\sigma^2$ normalized to the same ratio computed for $\mu_B=0$ and $T=T_c^0$ for three different values of the volume $\Omega$ as a function of the collision energy $\sqrt{s_{NN}}$, using its relation with $\mu_B$ given by Eq.~(\ref{therminatormus}). The upper panel is computed with $a=148.73$ MeV, $\lambda=1.4$ and $g=0.88$. The lower panel is computed with $a=141.38$ MeV, $\lambda=0.4$ and $g=0.82$. In each case the ratio $C_4/C_2$ is independent of $\Omega$ except near the collision energy where we find the CEP, and the high temperature approximation is less accurate. The value of $(\sqrt{s_{NN}})_\text{CEP}\sim 2$ GeV that corresponds to the CEP location for each set of parameters, is represented by the vertical line. The insets show the same ratio of cululants in a region around $(\sqrt{s_{NN}})_\text{CEP}$. Notice that $\kappa\sigma^2$ significantly drops down as the collision  energy moves from the right to the left across $(\sqrt{s_{NN}})_\text{CEP}$. This behavior is in agreement with recent HADES data~\cite{HADES:2020wpc}. }
    \label{kappasigmaplots}    
\end{figure}
\begin{figure}[t]
    \centering
    \includegraphics[scale=0.59]{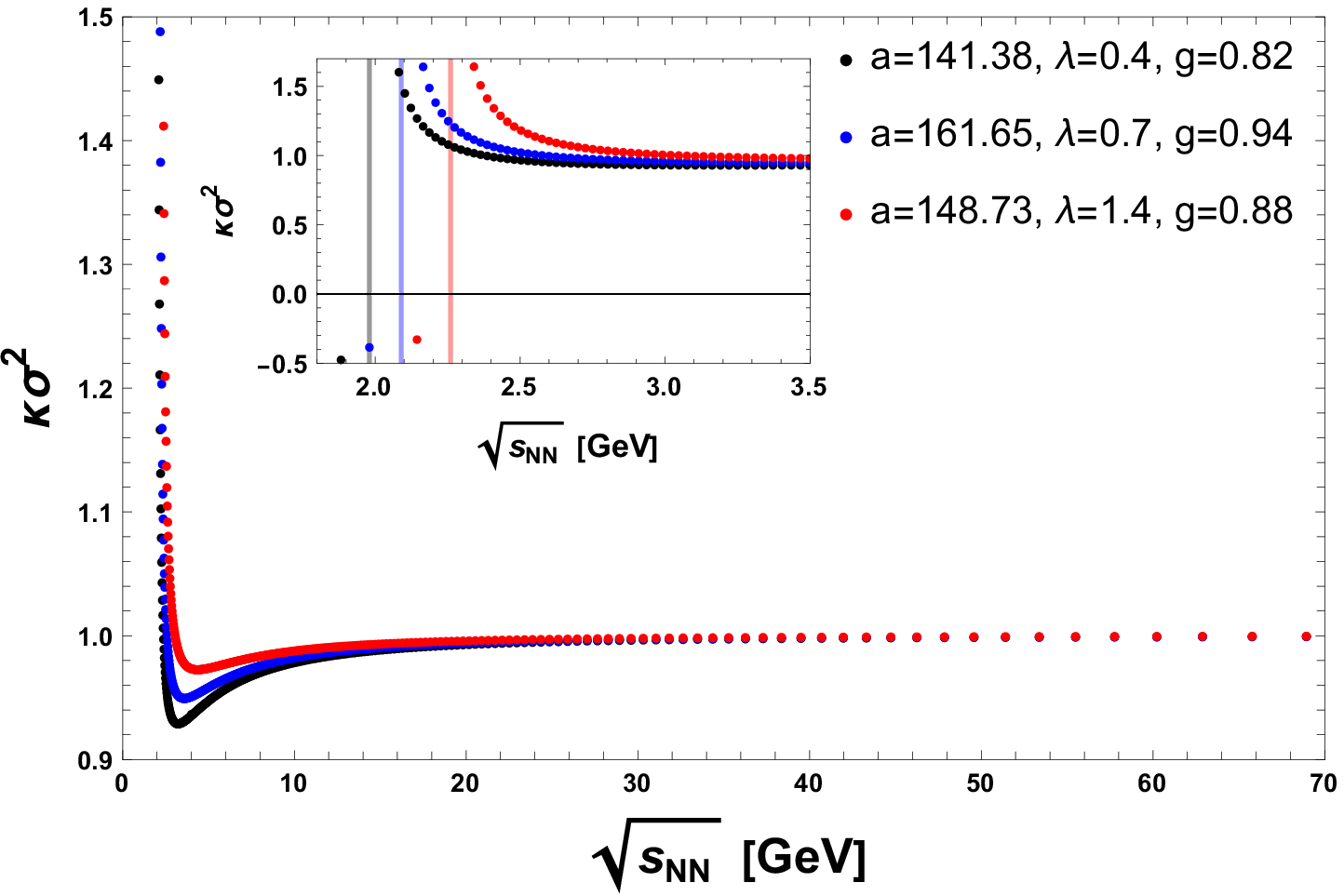}
    \caption{Ratio $C_4/C_2=\kappa\sigma^2$, normalized to the same ratio computed for $\mu_B=0$ and $T=T_c^0$ for $\Omega=(100\ \text{fm})^{3}$ for three different allowed sets of parameters $a$, $\lambda$ and $g$. The dips have different depths. However, the strong rise of the ratio happens for almost the same value of the collision energy $(\sqrt{s_{NN}})_\text{CEP}\sim 2$ GeV. This is shown in the inset where the same ratio of cululants in a region around $(\sqrt{s_{NN}})_\text{CEP}$ is depicted. Notice that $\kappa\sigma^2$ significantly drops down as the collision  energy moves from the right to the left across $(\sqrt{s_{NN}})_\text{CEP}$. This behavior is in agreement with recent HADES data~\cite{HADES:2020wpc}. }
    \label{kslambdas}
\end{figure}
To study the properties of the probability distribution we compute the behavior of its cumulants of third and fourth order. This is done in Fig.~\ref{sandk}. 
The upper panel shows the skewness $s$ whereas the lower panel shows the kurtosis $\kappa$, as functions of $\mu_B$, for fixed $T$ across the corresponding critical value of $\mu_B^c$, for $\lambda=0.4$, $g=0.82$ and $a=141.38$ MeV. Notice that for second order phase transitions, both $s$ and $\kappa$ are represented by smooth curves. However, when $T$ approaches $T^{\text{CEP}}$, these functions show a peaked structure that becomes more pronounced when the transitions become first order. Notice also that when $T$ goes below $T^{\text{CEP}}$, both cumulants develop a maximum for $\mu_B>\mu_B^c$.

We now turn to describing the behavior of these cumulants as functions of the collision energy in heavy-ion reactions. For this purpose, we resort to the relation between the chemical freeze-out value of $\mu_B$ and the collision energy $\sqrt{s_{NN}}$, given by~\cite{Cleymans:2005xv} 
\begin{eqnarray}
   \mu_B(\sqrt{s_{NN}})=\frac{d}{1+e\sqrt{s_{NN}}},
\label{therminatormus}
\end{eqnarray}
where $d=1.308$ GeV and $e=0.273$ GeV$^{-1}$. Figure~\ref{kappasigmaplots} shows the ratio of the cumulants $C_4/C_2=\kappa\sigma^2$, normalized to the same ratio computed for $\mu_B=0$ and $T=T_c$, where $\sigma^2$ is the variance, for three different values of the volume $\Omega$. The upper panel is computed with the set of parameters $a=148.73$ MeV, $\lambda=1.4$ and $g=0.88$, whereas the lower panel corresponds to $a=141.38$ MeV, $\lambda=0.4$ and $g=0.82$. The value of $\sqrt{s_{NN}}$ for each set of parameters that corresponds to the CEP location is represented by the vertical line. Thus, we see that the CEP position is heralded not by the dip of $C_4/C_2$ but for its strong rise as the energy that corresponds to the CEP is approached.A similar result has been found in Ref.~\cite{Mroczek:2020rpm}. Also, as pointed out in Ref.~\cite{Luo:2017uxa}, this non-monotonic
behavior cannot be described by many other model calculations~\cite{Xu:2016qjd,He:2016uei}. We emphasize that the reason for this failure is that other models do not incorporate long-range correlations and thus the possibility to describe critical phenomena, as we do in this work with the LSMq. Notice that in each case the ratio $C_4/C_2$ is independent of $\Omega$, as is expected for a ratio of cumulants, except near the collision energy where we find the CEP, where the high temperature approximation is less accurate. Nevertheless, the variations are of order of a few percent. Notice also that the product $\kappa\sigma^2$ significantly drops down for energies lower than the collision energy where the CEP is located. This result goes along recent  HADES measurements of net-proton fluctuations at low energies~\cite{HADES:2020wpc}.

Figure~\ref{kslambdas} shows the ratio $C_4/C_2=\kappa\sigma^2$, normalized to the same ratio computed for $\mu_B=0$ and $T=T_c^0$, for $\Omega=(200\ \text{fm})^{3}$ and for three different allowed sets of parameters $a$, $\lambda$ and $g$. Notice that although the dips have different depths, the strong rise of the ratio happens for almost the same value of the collision energy $\sqrt{s_{NN}}\sim 2$ GeV.

\section{Conclusions}\label{concl}

In conclusion, we have shown that the LSMq, with the effective potential computed up to the ring diagrams contribution and in the high temperature and chiral limits, produces an effective description of the system's equilibrium distribution that deviates from that of the HRGM, where the ratios of cumulants of even order are always equal to 1. The plasma screening properties, encoded in the contribution of the ring diagrams, make this deviation possible, since they describe a key feature of plasmas near phase transitions, namely, long-range correlations. We have fixed the LSMq parameters using conditions at the phase transition for $\mu_B=0$ provided by LQCD calculations, namely the crossover transition temperature $T_c^0$ and the curvature parameters $\kappa_2$ and $\kappa_4$. The phase diagram can be directly studied by finding the kind of phase transitions that the effective potential allows when varying $T$ and $\mu_B$. We found that the CEP can be located in the range 786 MeV $<\mu_B^{\text{CEP}}<849$ MeV and 69 MeV $< T^{\text{CEP}}<70.3$ MeV. From the probability distribution obtained using the effective potential, we have computed the behaviour of the kurtosis and of the skewness and found that these cumulants show strong peaks as the CEP is crossed. We have also computed the ratio of the cumulants $C_4/C_2=\kappa\sigma^2 $ as a function of the collision energy in a heavy-ion collision. We found that the location of the CEP coincides with the strong rise observed in this ratio and that this happens at $\sqrt{s_{NN}}\sim 2$ GeV. 

Our calculations have been carried out in the high temperature limit and without including an explicit symmetry breaking term that gives rise to a finite vacuum pion mass.
The high temperature approximation is valid near the (pseudo)transition lines, where what matters are the particles thermal masses. Since symmetry is being restored, the boson masses become small, of order $(\sqrt{\lambda/2 + g^2})T$, at small values of $\mu_B$, whereas for fermions they vanish given that at symmetry restoration $v=0$. Therefore, a high temperature expansion is a reasonable, although of limited accuracy, description scheme for temperatures $T \sim (\sqrt{\lambda/2 + g^2})T_c^0 \sim T_c^0$ and only down to the region where the contribution from $\mu_B$ produces that the thermal boson masses are not that small compared to $T$. 

The findings of this work are encouraging and can be extended to provide a more accurate description including an explicit chiral symmetry breaking introduced by a finite pion mass, as well as by going beyond the high temperature approximation. This is work for the near future that will be reported elsewhere. 

\section*{Acknowledgements}
Support for this work was received in part by UNAM-DGAPA-PAPIIT grant number IG100322 and by Consejo Nacional de Ciencia y Tecnolog\'ia grant numbers A1-S-7655 and A1-S-16215.  S. H.-O. acknowledges support from the U.S. DOE under Grant No. DE-FG02-00ER41132 and the Simons Foundation under the Multifarious Minds Program Grant No. 557037.

\bibliography{biblio}

\end{document}